\documentclass[aps,prd,superscriptaddress,nofootinbib,amsmath,amsfonts,preprintnumbers,groupedaddress,showpacs,10pt,english]{revtex4-1}
\usepackage{amsmath}
\usepackage{amssymb}
\usepackage{tabularx}
\usepackage{babel}
\usepackage{wrapfig}
\usepackage{float}
\usepackage{cancel}

\usepackage{relsize,exscale}
\usepackage{array}
\usepackage{float}
\makeatletter

\usepackage{array,multirow,graphicx}
\usepackage{soul}
\usepackage{dcolumn}
\usepackage{newlfont}
\usepackage{bm}
\usepackage[colorlinks,citecolor=blue,urlcolor=blue,linkcolor=blue]{hyperref}
\usepackage[figtopcap]{subfigure}
\usepackage{color}
\usepackage{verbatim}

\def\be{\begin{align}}
\def\ee{\end{align}}
\def\bea{\begin{eqnarray}}
\def\eea{\end{eqnarray}}
\def\bal{\begin{align}}
\def\eal{\end{align}}

\usepackage{scalerel}
\usepackage{tikz}
\usetikzlibrary{svg.path}
\definecolor{orcidlogocol}{HTML}{A6CE39}
\tikzset{
 orcidlogo/.pic={
 \fill[orcidlogocol] svg{M256,128c0,70.7-57.3,128-128,128C57.3,256,0,198.7,0,128C0,57.3,57.3,0,128,0C198.7,0,256,57.3,256,128z};
 \fill[white] svg{M86.3,186.2H70.9V79.1h15.4v48.4V186.2z}
 svg{M108.9,79.1h41.6c39.6,0,57,28.3,57,53.6c0,27.5-21.5,53.6-56.8,53.6h-41.8V79.1z M124.3,172.4h24.5c34.9,0,42.9-26.5,42.9-39.7c0-21.5-13.7-39.7-43.7-39.7h-23.7V172.4z}
 svg{M88.7,56.8c0,5.5-4.5,10.1-10.1,10.1c-5.6,0-10.1-4.6-10.1-10.1c0-5.6,4.5-10.1,10.1-10.1C84.2,46.7,88.7,51.3,88.7,56.8z};}}
\newcommand\orcid[1]{\href{https://orcid.org/#1}{\mbox{\scalerel*{
\begin{tikzpicture}[yscale=-1,transform shape]
\pic{orcidlogo};
\end{tikzpicture}
}{|}}}}
\graphicspath{{./}{Figs/}}
\begin{document}

\title{Regular Gravastar-Like Configurations in Quadratic Gauss--Bonnet Gravity}
\author{G.~G.~L.~Nashed} \email{nashed@bue.edu.eg}
\affiliation{Centre for Theoretical Physics, The British University in Egypt, P.O. Box 43, El Sherouk City, Cairo 11837, Egypt\\Centre for Space Research (CSR), North-West University, Potchefstroom 2520, South Africa}

\date{\today}

\begin{abstract}
We prescribe a smooth, static, spherically symmetric metric and use the
field equations of $f(R,\mathcal{G})=R+\beta\mathcal{G}^{2}$ gravity to
determine the density and the radial and tangential pressures of its
source. The metric is regular at the center and asymptotically flat after
a constant rescaling of time. A long-range radial pressure persists, so
the asymptotic region is nonvacuum. For $\bar\beta=0,0.5,1,3$, the radial
metric function stays positive on $0<r/L_0\leq20$, where the maximum of
$2m(r)/r$ is approximately $0.6866$, $0.7225$, $0.7584$, and $0.9030$,
respectively. At the three positive couplings, the density and the null
energy condition combinations become negative over finite radial
intervals, so the matter fails the global null and weak energy
conditions. The geometry has a regular center and a smooth anisotropic
transition. Defining a finite shell requires a separate prescription
for its boundaries. The Darmois--Israel formulas provide the GR
reference as $\beta\to0$. A stability calculation at nonzero $\beta$
requires the junction conditions of the quadratic theory and an
independent law for the surface stresses.
\end{abstract}
\maketitle
\section{Introduction}

General relativity predicts black-hole formation when sufficiently
compact matter collapses \cite{Hawking:1973uf,Misner:1973prb}.
Black-hole metrics describe observed compact objects well. The
classical singularity and the information-loss problem motivate the study
of regular alternatives \cite{Penrose:1965rx,Hawking:1975vcx}.
The Mazur--Mottola gravastar places a thin shell between a de Sitter
interior and a Schwarzschild exterior \cite{Mazur:2004fk}. Continuous
gravastar models use anisotropic stresses to connect the central and outer
regions \cite{Visser:2003ge,Cattoen:2005he}. Their possible observational
differences from black holes have also been examined \cite{Chirenti:2007mk}.
Wormholes and anisotropic stars provide other examples of regular compact
geometries \cite{Lobo:2009ip,Rahaman:2013xoa}.

On the other hand, general relativity may break down in regions of very high curvature, which has motivated the study of modified gravity theories in both cosmology and strong-field physics \cite{Nojiri:2010wj,Mofarreh:2026xjh,Clifton:2011jh,Nashed:2024pbc,Capozziello:2011et}. Among these, $f(R,\mathcal{G})$ gravity is especially interesting because its Lagrangian depends on both the Ricci scalar $R$ and the Gauss--Bonnet invariant $\mathcal{G}$ \cite{Nojiri:2005jg,DeFelice:2010aj}. In four dimensions, the term linear in $\mathcal{G}$ is purely topological and does not affect the local field equations, but nonlinear functions of $\mathcal{G}$ can contribute nontrivially to the dynamics. Higher-curvature terms of this kind are also well motivated by string-inspired effective gravity theories \cite{Gross:1986mw,Zwiebach:1985uq}, where they arise as natural corrections and may become important in strong-curvature regions, altering the structure of compact objects. Indeed, modified-gravity studies of stars and wormholes show that such corrections can significantly change the effective density, pressures, anisotropy, and physical viability conditions compared with general relativity \cite{Astashenok:2013vza,Zubair:2016cde,Shamir:2018qhq,Nashed:2021ctg,Boehmer:2011gw,Nashed:2023uvk,Sharif:2020xxj}. This makes it natural to examine how a quadratic Gauss--Bonnet correction affects a regular gravastar-like geometry.

Bhatti, Yousaf, and Rehman constructed a three-region gravastar in
$f(R,\mathcal G)$ gravity using prescribed equations of state and Israel
matching \cite{Bhatti:2020cjz}. Bhatti, Yousaf, and Ashraf calculated the
length, energy, and entropy of a shell in a related modified Gauss--Bonnet
model \cite{Bhatti:2021dvp}. In Einstein gravity, constrained scalar fields
have been used to connect a de Sitter interior to a Schwarzschild exterior
through a finite, smooth shell, with an analysis of surface redshift
and stability \cite{Nojiri:2023zlp}.

Here we specify one smooth metric depending on $\beta$ and use the
full $R+\beta\mathcal G^2$ field equations to obtain $\rho(r)$, $p_r(r)$,
and $p_t(r)$. This calculation identifies the radial intervals that
violate the null and weak energy conditions. The large-$r$ expansion
gives a $1/r$ term in the radial metric and a leading correction of
order $1/r^2$ in the temporal metric. These falloffs leave a radial
pressure tail. We use the term gravastar-like for the regular central
region and the smooth anisotropic transition. A three-region model
would require physical shell boundaries and a separate vacuum
exterior. We discuss the GR junction formulas and the dynamical
input needed for a thin-shell stability calculation.

 The rest of the paper is organized as follows: the next section introduces the basic equations of quadratic $f(R,\mathcal{G})$ gravity; Sec.\ref{IV} analyzes the central, transition, and asymptotic regions; Sec.\ref{VI} and \ref{VII} examine the physical properties and energy conditions; Sec.\ref{IX} studies the effect of the coupling parameter $\beta$; and the final section summarizes the results and discussion.
\section{Formalism of $f(R,\mathcal{G})$ Gravity}\label{II}


In modified theories of gravity, the Einstein--Hilbert action is generalized by including higher-order curvature invariants in order to account for quantum corrections, cosmological acceleration, and deviations from General Relativity (GR) at high energies \cite{Nojiri:2010wj,Asghar:2026que, Clifton:2011jh}. One such extension is the $f(R,\mathcal{G})$ gravity, where the gravitational Lagrangian depends on both the Ricci scalar $R$ and the Gauss--Bonnet invariant $\mathcal{G}$.

In this work, we consider the specific quadratic Gauss--Bonnet model given by
\begin{equation}\label{lag}
S = \frac{1}{2\kappa^2} \int d^4x \sqrt{-g} \left(R + \beta \mathcal{G}^2\right) + S_m,
\end{equation}
where $\kappa^2 = 8\pi G$ (we adopt geometrized units with $G=c=1$ unless stated otherwise), $g$ is the determinant of the metric tensor $g_{\mu\nu}$, and $S_m$ represents the matter action. The Gauss--Bonnet invariant $G$ is defined as
\begin{equation}
\mathcal{G} = R^2 - 4R_{\mu\nu}R^{\mu\nu} + R_{\mu\nu\rho\sigma}R^{\mu\nu\rho\sigma},
\end{equation}
where $R_{\mu\nu}$ and $R_{\mu\nu\rho\sigma}$ denote the Ricci and Riemann tensors, respectively.

In four dimensions the integral of the linear Gauss--Bonnet invariant
is topological and gives no bulk contribution to the metric field
equations \cite{Lovelock:1971yv}. The term $\mathcal G^2$ contributes
algebraic curvature terms as well as derivatives of $\mathcal G$
\cite{Nojiri:2005jg,DeFelice:2010aj}. We write the matter action as
\begin{equation}
S_m = \int d^4x \sqrt{-g}\,\mathcal{L}_m,
\end{equation}
where $\mathcal L_m$ is the matter Lagrangian density. Reviews of
higher-curvature gravity and its applications include
Refs.~\cite{Nojiri:2010wj, Capozziello:2011et}.

In Eq.~\eqref{lag}, $R$ has units of $L^{-2}$ and $\mathcal G$ has
units of $L^{-4}$, so $\beta$ has units of $L^6$. We vary $\beta$
as a free coupling and use $\beta=0$ as the GR reference.
For background on curvature terms in gravitational actions, see
Refs.~\cite{Gross:1986mw, Zwiebach:1985uq, Nojiri:2010wj, Capozziello:2011et}.


The field equations of the theory are obtained by varying the action with respect to the metric tensor $g_{\mu\nu}$. In the framework of $f(R,\mathcal{G})$ gravity, this variation leads to modified Einstein equations containing additional contributions from the higher-order curvature invariants \cite{Nojiri:2005jg,DeFelice:2010aj}. For a general $f(R,\mathcal{G})$ theory, it is convenient to define
\begin{equation}
f_R \equiv \frac{\partial f}{\partial R}, \qquad
f_{\mathcal{G}} \equiv \frac{\partial f}{\partial \mathcal{G}}.
\end{equation}
For the specific model considered here,
\begin{equation}
f(R,\mathcal{G}) = R + \beta \mathcal{G}^2,\qquad \mbox{we obtain} \qquad
f_R = 1, \qquad f_{\mathcal{G}} = 2\beta \mathcal{G}.
\end{equation}

The variation of the action given by Eq.~\eqref{lag} yields the field equations in the form
\begin{equation}
G_{\mu\nu} + \beta \mathcal{H}_{\mu\nu} = \kappa^2 T_{\mu\nu},
\end{equation}
where $G_{\mu\nu} = R_{\mu\nu} - \frac{1}{2} R g_{\mu\nu}$ is the Einstein tensor, $T_{\mu\nu}$ is the energy-momentum tensor derived from the matter Lagrangian, and $\mathcal{H}_{\mu\nu}$ encodes the modifications arising from the quadratic Gauss--Bonnet term \cite{Nojiri:2024cal,Nojiri:2023qgd}.

In the convention used here, $T_{\mu\nu}$ is the matter tensor required by the prescribed metric through $\kappa^2T_{\mu\nu}=G_{\mu\nu}+\beta H_{\mu\nu}$. It should not be confused with the alternative ``effective curvature fluid'' obtained by moving $H_{\mu\nu}$ to the right-hand side. Energy conditions below refer specifically to this required matter tensor.

We use the curvature convention
\begin{equation}
R^{\rho}{}_{\sigma\mu\nu}=\partial_{\mu}\Gamma^{\rho}_{\nu\sigma}
-\partial_{\nu}\Gamma^{\rho}_{\mu\sigma}
+\Gamma^{\rho}_{\mu\lambda}\Gamma^{\lambda}_{\nu\sigma}
-\Gamma^{\rho}_{\nu\lambda}\Gamma^{\lambda}_{\mu\sigma}.
\label{riemann_convention}
\end{equation}
For this convention, the tensor $\mathcal{H}_{\mu\nu}$ obtained from the variation of $\mathcal G^2$ is
\begin{align}
H_{\mu\nu}={}&
4\mathcal{G}R R_{\mu\nu}
-8\mathcal{G}R_{\mu\alpha}R^{\alpha}{}_{\nu}
-8\mathcal{G}R_{\mu\alpha\nu\beta}R^{\alpha\beta}
+4\mathcal{G}R_{\mu}{}^{\alpha\beta\gamma}
R_{\nu\alpha\beta\gamma}
-\frac{1}{2}g_{\mu\nu}\mathcal{G}^{2}
\nonumber\\
&-4R\nabla_{\mu}\nabla_{\nu}\mathcal{G}
+4g_{\mu\nu}R\Box\mathcal{G}
+8R^{\alpha}{}_{\nu}\nabla_{\alpha}\nabla_{\mu}\mathcal{G}
+8R^{\alpha}{}_{\mu}\nabla_{\alpha}\nabla_{\nu}\mathcal{G}
\nonumber\\
&-8R_{\mu\nu}\Box\mathcal{G}
-8g_{\mu\nu}R^{\alpha\beta}
\nabla_{\alpha}\nabla_{\beta}\mathcal{G}
+8R_{\mu\alpha\nu\beta}
\nabla^{\alpha}\nabla^{\beta}\mathcal{G}.
\end{align}
where $\nabla_\mu$ denotes the covariant derivative and $\Box \equiv \nabla^\alpha \nabla_\alpha$ is the d'Alembert operator.

Since $\mathcal G$ is built from curvature, the terms containing
$\nabla_\mu\nabla_\nu\mathcal G$ can bring fourth derivatives of the
metric into the field equations \cite{Nojiri:2010wj, Capozziello:2011et}.

The energy-momentum tensor is defined in the standard way as
\begin{equation}
T_{\mu\nu} = -\frac{2}{\sqrt{-g}} \frac{\delta (\sqrt{-g}\mathcal{L}_m)}{\delta g^{\mu\nu}}.
\end{equation}
Due to the diffeomorphism invariance of the total action, the field equations satisfy the conservation condition
\begin{equation}
\nabla^\mu T_{\mu\nu} = 0,\quad \mbox{which ensures the local conservation of energy and momentum.}
\end{equation}
Finally, we should emphasize that in the limit $\beta \rightarrow 0$, the modified field equations reduce to those of General Relativity\footnote{Throughout this study, we adopt relativistic units in which $c=G=1$ (and $\hbar=1$ where applicable).},
\begin{equation}
G_{\mu\nu} = \kappa^2 T_{\mu\nu},
\end{equation}
thereby recovering the standard theory as a special case. This gives an important consistency check of the model.

For the static, spherically symmetric gravastar-like configuration considered here, we specialize the general field equations to the metric introduced in the following section. The \(tt\), \(rr\), and \(\theta\theta\) components of the modified field equations then determine the required matter energy density \(\rho(r)\), radial pressure \(p_r(r)\), and tangential pressure \(p_t(r)\) associated with the chosen metric ansatz.

\section{Metric and Matter Content}\label{III}

In the present study, we consider a static and spherically symmetric spacetime. Such a symmetry is the natural first approximation for compact objects and has been widely used in the study of stars, black holes, and gravastars \cite{Mazur:2004fk, Visser1995}. The line element is given by
\begin{align}\label{met}
ds^2 = -e^{a(r)} dt^2 + e^{b(r)} dr^2 + r^2 d\Omega^2,\quad \text{where} \quad
d\Omega^2 = d\theta^2 + \sin^2\theta\, d\phi^2,
\end{align}
with $d\Omega^2$ representing the metric on the unit two-sphere.

For the regular gravastar-like geometry, we choose the following metric potentials
\begin{align}
a(r)= c_0 - c_1 \frac{r^2}{R^2 + r^2} + \beta c_2 \frac{r^4}{(R^2 + r^2)^2},
\qquad
e^{-b(r)} = 1-\frac{2M}{r}\left(1-e^{-r^3/R^3}\right)-\beta\,\frac{2c_3r^2}{(R^2+r^2)^2}.
\label{ansatz_a}
\end{align}
{ The ansatz in Eq.~(\ref{ansatz_a}) is chosen as a smooth geometry-driven
interpolation between a regular center and an asymptotically flat
region. The factor \(1-e^{-r^{3}/R^{3}}\) gives
\(m(r)\sim r^{3}\) near \(r=0\), ensuring \(e^{-b(0)}=1\), while
\(m(r)\to M\) asymptotically, producing the Schwarzschild-type
\(1/r\) falloff of the radial metric component. The bounded rational
terms in \(a(r)\) ensure \(a'(0)=0\), and the \(\beta\)-dependent
terms introduce smooth higher-curvature deformations without spoiling
central regularity or the leading asymptotic behavior. This ansatz is
phenomenological and not unique; therefore, the reconstructed matter
profiles and energy-condition results are conditional on this metric
choice.}

The radial metric function \(f(r)\equiv e^{-b(r)}\) can be written in terms of an effective mass function \(m(r)\), defined through \(f(r)=1-2m(r)/r\). The chosen form ensures regularity at the center, where \(m(r)\sim r^3\), and a Schwarzschild-like
\(1/r\) falloff in the radial metric component, since
\[
e^{-b(r)}\longrightarrow 1-\frac{2M}{r}
\qquad \text{as}\qquad r\longrightarrow\infty.
\]
The term proportional to \(\beta\) in the radial function enters at
order \(1/r^2\). The first correction to the constant temporal
potential is also of order \(1/r^2\). Section~\ref{IV} uses these
falloffs to obtain the asymptotic radial pressure.

Here $c_0$, $c_1$, $c_2$, $c_3$, $R$, and $M$ are constants, while $\beta$ is the quadratic Gauss--Bonnet coupling parameter appearing in the gravitational Lagrangian $f(R,\mathcal{G})=R+\beta \mathcal{G}^2$. Dimensional consistency requires
\begin{equation}
[c_0]=[c_1]=1,\qquad [c_2]=L^{-6},\qquad [c_3]=L^{-4},
\qquad [M]=[R]=L,\qquad [\beta]=L^6.
\end{equation}
For the numerical analysis, it is convenient to introduce a reference length $L_0$ and define
\begin{equation}
x=\frac r{L_0},\quad \bar M=\frac M{L_0},\quad
\bar R=\frac R{L_0},\quad \bar\beta=\frac\beta{L_0^6},\quad
\bar c_2=c_2L_0^6,\quad \bar c_3=c_3L_0^4.
\label{dimensionless_parameters}
\end{equation}
All numerical values quoted below refer to these dimensionless quantities, even where bars on $M,R,c_2,c_3$ are suppressed to avoid clutter.

We should note that the above ansatz is regular at the center $r=0$. In fact,
\begin{align}
e^{-b(0)}=1, \qquad a'(0) = 0,
\end{align}
Together with the smooth near-center expansions of the metric functions, these conditions ensure a regular center with finite curvature invariants. The $\beta$-dependent terms encode corrections induced by the quadratic Gauss--Bonnet invariant and become important in regions of high curvature, particularly in the transition region surrounding the regular central core.

We assume that the matter sector is described by an anisotropic fluid. This choice is physically motivated, as anisotropy naturally arises in highly compact objects due to strong gravitational fields and high-density effects \cite{Bowers:1974tgi, Herrera:1997plx}. The energy-momentum tensor is written as
\begin{equation}
T^\mu_{\ \nu} = \mathrm{diag}(-\rho\,, p_r\,, p_t\,, p_t),
\label{Tmunu_mixed}
\end{equation}
where $\rho=\rho(r)$, $p_r=p_r(r)$, and $p_t=p_t(r)$ denote the energy density, radial and tangential pressures, respectively.

In covariant form,
\begin{equation}
T_{\mu\nu} = (\rho+p_t)u_\mu u_\nu + p_t g_{\mu\nu} + (p_r-p_t)\chi_\mu \chi_\nu,
\label{Tmunu_cov}
\end{equation}
where $u^\mu$ is the four-velocity and $\chi^\mu$ is a unit radial spacelike vector satisfying
\begin{align}
u^\mu u_\mu = -1,\qquad \chi^\mu \chi_\mu = 1,\qquad u^\mu \chi_\mu = 0.
\end{align}
For the present metric,
\begin{align}
u^\mu = e^{-a(r)/2}\delta^\mu_0, \qquad \chi^\mu = e^{-b(r)/2}\delta^\mu_1,\quad \mbox{leading to} \quad
T_{tt} = \rho\, e^{a(r)}, \quad
T_{rr}= p_r\, e^{b(r)}, \quad
T_{\theta\theta} = p_t\, r^2, \quad
T_{\phi\phi} = p_t\, r^2 \sin^2\theta.
\end{align}

For the metric in Eq.~\eqref{met}, the curvature scalar $R$ and the Gauss--Bonnet invariant $\mathcal{G}$ are determined by the metric functions and their radial derivatives. For the gauge in Eq.~\eqref{met}, the reduced expressions may be written in terms of $a$, $f=e^{-b}$, $a'$, $a''$, and $f'$; no independent second derivative of $b$ is required.
Using the metric ansatz (\ref{ansatz_a}) in the modified field equations, we obtain the equations which determine the matter variables from the geometry. In particular, the $tt$, $rr$, and $\theta\theta$ components determine $\rho(r)$, $p_r(r)$, and $p_t(r)$ in terms of $a(r)$ and $b(r)$ and their derivatives. Appendix~\ref{app:H} records the covariant higher-curvature tensor and Appendix~A gives analytic near-center checks. The full reduced component expressions are lengthy, so the complete covariant derivation, symbolic benchmark tests, numerical evaluation, zero finding, conservation diagnostic, and figure generation are supplied in the accompanying Python program.

We define the anisotropy of the fluid by
\begin{equation}
\Delta = p_t - p_r, \quad \mbox{which plays an important role in the equilibrium and stability of the configuration.}
\end{equation}
The conservation equation $\nabla^\mu T_{\mu\nu}=0$ leads to the generalized Tolman--Oppenheimer--Volkoff equation
\begin{equation}
\frac{dp_r}{dr} = -\frac{1}{2}(\rho+p_r)\frac{da}{dr} + \frac{2}{r}(p_t-p_r), \quad \mbox{which governs hydrostatic equilibrium.}
\end{equation}

In order to examine the physical admissibility of the solution, we use the following diagnostics: regularity of the metric and matter variables at the center, Lorentzian signature with $f(r)>0$ on the static domain, positivity of the matter density where required, and the relevant energy conditions. They are tests rather than assumptions: as shown below, the representative positive-$\bar\beta$ examples fail some of the matter conditions over finite radial intervals. Moreover, because the present ansatz has a nonvacuum long-range pressure tail, it should not be interpreted as a complete isolated compact object without an additional matching prescription.

\section{Gravastar--Like Structure}\label{IV}

For the parameters used below, the center has negative pressure and the
leading radial metric function has de Sitter form. The radial and
tangential pressures separate away from the center, and the geometry
passes smoothly into a nonvacuum tail. These radial regimes motivate
the term gravastar-like. We examine the central equation of state below.
A three-region construction would also require finite boundaries and
a separate vacuum exterior. Section~\ref{V} gives the GR formulas
that would enter the corresponding matching problem.


The interior region is characterized by the behavior of the metric functions near the center $r=0$. Expanding the ansatz functions around the origin, we obtain
\begin{align}
a(r) &= c_0 -c_1 \frac{r^2}{R^2} + \mathcal{O}(r^4),
\end{align}
where this expansion follows from the bounded form $r^2/(R^2+r^2)$ in the metric potential. For the radial component, we have
\begin{align}
e^{-b(r)}
&= 1-\frac{2M}{r}\left(1-e^{-r^3/R^3}\right)-\beta\frac{2c_3r^2}{(R^2+r^2)^2} \nonumber\\
&=1-\left(\frac{2M}{R^3}+\frac{2\beta c_3}{R^4}\right)r^2
+\frac{4\beta c_3}{R^6}r^4+\frac{M}{R^6}r^5+\mathcal{O}(r^6).
\end{align}

Thus, to leading order,
\begin{equation}
e^{-b(r)} \approx 1-\frac{r^2}{\ell_{\mathrm{eff}}^2},\qquad \mbox{where}\qquad
\frac{1}{\ell_{\mathrm{eff}}^2} = \frac{2M}{R^3}+\frac{2\beta c_3}{R^4}.
\end{equation}

We find from this behavior that the spacetime is regular at the origin. The metric functions admit smooth near-center expansions, with $e^{-b(0)}=1$ and $a'(0)=0$; together these conditions give a regular areal-radius center with finite curvature invariants. The leading-order structure of the metric is therefore analogous to a de Sitter spacetime.

Regularity implies finite central values and central isotropy,
\begin{equation}
\rho(r)=\rho_0+\mathcal O(r^2),\qquad
p_r(0)=p_t(0),
\end{equation}
but it does not imply the de Sitter equation of state $p(0)=-\rho(0)$. In the GR limit the present expansion gives
\begin{equation}
\kappa^2\rho(0)=3A,\qquad
\kappa^2p_r(0)=\kappa^2p_t(0)=2\alpha-A,
\end{equation}
with $A=2M/R^3$ and $\alpha=-c_1/R^2$. For the representative GR parameters used below, $A=1/4$ and $\alpha=-1/5$, so
\begin{equation}
\frac{p(0)}{\rho(0)}=-\frac{13}{15}\simeq-0.867,
\end{equation}
rather than $-1$. For $\beta\ne0$ the finite $H^\mu{}_{\nu}(0)$ terms must be evaluated explicitly. Therefore, in the present paper we regard the central region as regular and gravastar-like, but not as an exact de Sitter core.


If an independent prescription later identifies radii $r_1<r_2$, the interval $r_1\leq r\leq r_2$ could be interpreted as an effective shell. In the present smooth solution we instead speak of a transition region, where the metric gradients and higher-curvature corrections may become appreciable.

In the usual gravastar construction, the intermediate shell is often
modeled by an ultra-relativistic stiff equation of state,
\begin{equation}
p_r(r)\approx \rho(r).
\end{equation}
Here the field equations determine the matter profiles. We use
$p_r/\rho$ in Fig.~\ref{fig:eos} to test how closely the reconstructed
matter approaches the stiff-fluid value.

The anisotropy factor
\begin{equation}
\Delta(r) = p_t(r) - p_r(r),
\end{equation}
enters the radial force balance as $2\Delta/r$. This term acts outward
where $\Delta>0$ and inward where $\Delta<0$. The transition profiles
depend on $R$, $M$, and $\beta$ through the metric and its curvature.
Assigning a physical thickness to the transition requires two boundary
radii fixed by an independent criterion.

The large-radius behavior follows from Eq.~\eqref{ansatz_a}. For the
radial component we obtain
\begin{equation}
e^{-b(r)} = 1-\frac{2M}{r}-\frac{2\beta c_3}{r^2}+\mathcal{O}\left(\frac{1}{r^4}\right),
\end{equation}
whereas
\begin{equation}
a(r)=a_\infty+\frac{R^2(c_1-2\beta c_2)}{r^2}+\mathcal O(r^{-4}),
\qquad a_\infty=c_0-c_1+\beta c_2.
\end{equation}
After a constant rescaling of $t$, $g_{tt}\to-1$, but it has no Schwarzschild $1/r$ term. Indeed, already in the GR limit,
\begin{equation}
G^r{}_{r}=\frac{f-1}{r^2}+\frac{fa'}r
=-\frac{2M}{r^3}+\mathcal O(r^{-4}),
\end{equation}
so the required matter has a long-range pressure tail. The exterior is therefore neither an exact vacuum nor a Schwarzschild region.


We use core and shell as qualitative descriptions of the central and
transition regimes. Finite boundaries would require a criterion such
as a tolerance on the equation-of-state ratio, together with matching
to a separate exterior. The profiles at different $\beta$ include both
the change in the prescribed metric and the change in the
Gauss--Bonnet contribution to its source.
\begin{figure}[H]
\centering
\includegraphics[width=0.3\textwidth]{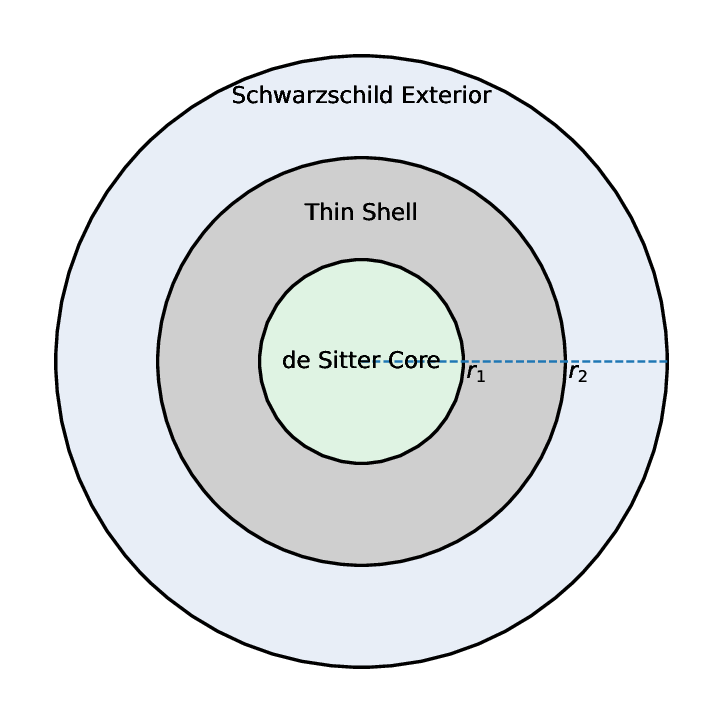}
\caption{Reference schematic of the standard gravastar idealization. The smooth ansatz studied here has a regular, not exactly de Sitter, center and an asymptotically flat, nonvacuum exterior; the three boundaries shown in this schematic are not derived from the ansatz.}
\label{fig:gravastar_structure}
\end{figure}

\section{Junction Conditions}\label{V}
For completeness, we now recall the standard Darmois--Israel junction conditions \cite{Israel:1966rt,Nashed:2021gkp}. A putative effective thin shell is placed at $r=R_\Sigma$, where $R_\Sigma$ is distinct from the scale $R$ in Eq.~\eqref{ansatz_a}. The present smooth ansatz does not determine $R_\Sigma$ or the two geometries to be joined.

The matching is then carried out using the Darmois--Israel formalism, which relates the discontinuity in the extrinsic curvature across the hypersurface to the surface stress-energy tensor localized on the shell.

We consider a timelike hypersurface $\Sigma$ defined by
\begin{equation}
r = R_\Sigma = \text{constant}.
\end{equation}
Using proper time $\tau$ on a static shell, the induced metric is
\begin{equation}
ds^2_{\Sigma}=-d\tau^2+R_\Sigma^2d\Omega^2.
\end{equation}
The first junction condition is continuity of this first fundamental form, $[h_{ij}]=0$. It identifies the same areal radius on both sides. The coordinate times $t_+$ and $t_-$ may be normalized independently, with $d\tau=e^{a_\pm(R_\Sigma)/2}dt_\pm$ for a static shell; therefore equality of the numerical lapse values $e^{a_+}$ and $e^{a_-}$ is not, by itself, a coordinate-invariant matching requirement.


The extrinsic curvature is defined as
\begin{equation}
K_{ij} = -n_\mu \left( \frac{\partial^2 x^\mu}{\partial \xi^i \partial \xi^j}
+\Gamma^\mu_{\alpha\beta} \frac{\partial x^\alpha}{\partial \xi^i}
  \frac{\partial x^\beta}{\partial \xi^j} \right),\qquad
  \mbox{with unit normal $n^\mu$.}
  \end{equation}
 For the metric under consideration, the outward-pointing unit normal vector is
\begin{equation}
n_\mu = (0,\, e^{b(r)/2},\, 0,\, 0).
\end{equation}
The non-vanishing components of the extrinsic curvature are:

\paragraph*{Angular components}
\begin{align}
K_{\theta\theta} = R_\Sigma e^{-b(R_\Sigma)/2}, \qquad
K_{\phi\phi} = R_\Sigma e^{-b(R_\Sigma)/2} \sin^2\theta.
\end{align}

\paragraph*{Temporal component}
\begin{equation}
K_{tt} = -\frac{1}{2} a'(R_\Sigma) e^{a(R_\Sigma)} e^{-b(R_\Sigma)/2},
\end{equation}
where the derivative is taken in the bulk metric on the side being evaluated.
Raising the first index gives the following mixed components and trace.
\begin{equation}
K^\theta_{\ \theta} = \frac{e^{-b(R_\Sigma)/2}}{R_\Sigma}, \qquad
K^t_{\ t} = \frac{1}{2} a'(R_\Sigma) e^{-b(R_\Sigma)/2}\Rightarrow
K = \frac{1}{2} a'(R_\Sigma) e^{-b(R_\Sigma)/2} + \frac{2}{R_\Sigma} e^{-b(R_\Sigma)/2}.
\end{equation}
The surface stress-energy tensor is given by
\begin{equation}
S^i_j = -\frac{1}{8\pi} \left( [K^i_j] - \delta^i_j [K] \right),\qquad \mbox{where $[X] = X_+ - X_-$ \,\,\,denotes the jump across the hypersurface.}
\end{equation}

For a spherically symmetric shell,
\begin{equation}
S^i_j = \text{diag}(-\sigma, \mathcal{P}, \mathcal{P}), \qquad \mbox{where $\sigma$ is the surface energy density and $\mathcal{P}$ is the surface pressure.}
\end{equation}
Using the extrinsic curvature components, we obtain:

\paragraph*{Surface energy density}
\begin{equation}\label{39}
\sigma = -\frac{1}{4\pi R_\Sigma} \left( e^{-b_+(R_\Sigma)/2} - e^{-b_-(R_\Sigma)/2} \right).
\end{equation}

\paragraph*{Surface pressure}
\begin{equation}\label{40}
\mathcal{P} = \frac{1}{8\pi} \left[
\frac{a'_+(R_\Sigma)}{2} e^{-b_+(R_\Sigma)/2}
-\frac{a'_-(R_\Sigma)}{2} e^{-b_-(R_\Sigma)/2}+ \frac{1}{R_\Sigma} \left( e^{-b_+(R_\Sigma)/2} - e^{-b_-(R_\Sigma)/2} \right)
  \right].
  \end{equation}


In terms of $f(r)=e^{-b(r)}$, the surface pressure in Eq.~\eqref{40}
uses $a'(R_\Sigma)\sqrt{f(R_\Sigma)}$ on each side of the surface.
\paragraph*{Matching in the quadratic theory.}
{
Equations~(43)--(44) are the standard GR Darmois--Israel expressions
and cannot be assumed to be the complete distributional junction
conditions of the nonlinear \(R+\beta\mathcal G^2\) theory. A consistent
matching would first require a finite radius \(R_\Sigma\) and a
separately constructed asymptotically flat vacuum exterior satisfying
the same modified field equations; the Schwarzschild geometry cannot
be imposed a priori. The induced metric must be continuous across
\(\Sigma\). Because the field equations contain derivatives of the
curvature, a regular matching without a surface layer must also exclude
distributional singularities generated by jumps in the relevant curvature
data. A deliberately stronger, conservative sufficient regularity
prescription is
\[
 [h_{ij}]=0,\qquad [K_{ij}]=0,\qquad
 [R_{\alpha\beta\gamma\delta}]=0,\qquad
 [n^\mu\nabla_\mu R_{\alpha\beta\gamma\delta}]=0 .
\]
These conditions are stated here as sufficient regularity requirements,
not as a derivation of the most general junction conditions of the
\(\mathcal G^2\) theory. In a formulation adapted to the Gauss--Bonnet sector,
one must in particular control possible jumps of \(\mathcal G\) and of
the normal derivative of \(f_{\mathcal G}=2\beta\mathcal G\). If a
genuine thin shell is allowed, its surface
stress tensor must instead be derived directly from the distributional
field equations of the \(\mathcal G^2\) theory and cannot be identified with
Eqs.~(43)--(44) alone.

The present ansatz does not specify distinct interior and exterior
solutions, a physical matching radius, or the required higher-curvature
surface data. We therefore retain Eqs.~(43)--(44) only as GR-limit
diagnostics and make no claim that a consistent thin-shell matching
has already been constructed. Completing such a construction requires
solving the exterior vacuum equations and applying the junction
conditions specialized to \(R+\beta\mathcal G^2\) gravity.}

{The field equations and conservation identity test static equilibrium.
A radial perturbation calculation requires a moving surface
$R_\Sigma(\tau)$, junction conditions for $R+\beta\mathcal G^2$, and
an evolution law for the surface stresses. Once these equations close,
a potential-based test can ask whether a static radius satisfies
\[
V(R_0)=V'(R_0)=0,\qquad V''(R_0)>0
\]
on the resulting dynamical branch. The present calculation checks
regularity and the absence of horizons on the stated radial interval.
Linear radial stability remains open.}

\section{Physical Properties}\label{VI}
In the units of Eq.~\eqref{dimensionless_parameters}, we use $c_0=0.5$,
$c_1=0.8$, $\bar c_2=0.8$, $\bar c_3=0.6$, $\bar M=1$, and $\bar R=2$,
with $\bar\beta=0,0.5,1,3$. Central regularity leaves $c_1$ free.
This coefficient enters the central pressure through the quadratic
term in the lapse potential, as shown in Sec.~\ref{IV}.

We first examine the condition $f(r)>0$ for the displayed family. A direct numerical minimization over $0<r/L_0\leq20$ gives
\begin{center}
\begin{tabular}{c|cccc}
$\bar\beta$ & $0$ & $0.5$ & $1$ & $3$\\ \hline
$\min f$ & $0.3134$ & $0.2775$ & $0.2416$ & $0.0970$
\end{tabular}
\end{center}
Thus no zero of the radial metric function occurs on this domain. This horizon check does not establish matter viability.

The gap between the static branch and $f=0$ can be read from the local
compactness. With the Misner--Sharp-type mass defined by $f(r)=1-2m(r)/r$,
we obtain
\begin{equation}
\mathcal C(r)\equiv\frac{2m(r)}{r}=1-f(r).
\label{compactness_diagnostic}
\end{equation}
Here $r$ is the areal radius, so $\mathcal C(r)$ is a geometric measure
at each radius. The surface compactness $2M/R_\star$ of a star would
require a separately defined surface $R_\star$. On the same numerical
interval, the maxima $1-\min f$ are
\begin{center}
\begin{tabular}{c|cccc}
$\bar\beta$ & $0$ & $0.5$ & $1$ & $3$\\ \hline
$\max\mathcal C$ & $0.6866$ & $0.7225$ & $0.7584$ & $0.9030$
\end{tabular}
\end{center}
for the representative family. Thus the positive-$\bar\beta$ members become progressively more compact in this geometrical sense while remaining horizon-free on the inspected domain. Since $\beta$ also occurs explicitly in the ansatz, this trend compares different geometries as well as different theories and should not be interpreted as a universal response at fixed geometry.

For the numerical plots we use the dimensionless matter variables
\begin{equation}
\bar\rho=\kappa^2L_0^2\rho,\qquad
\bar p_r=\kappa^2L_0^2p_r,\qquad
\bar p_t=\kappa^2L_0^2p_t,
\label{barred_matter}
\end{equation}
and $x=r/L_0$. The Python implementation constructs the four-dimensional curvature tensors covariantly using Eq.~\eqref{riemann_convention}, and then specializes to Eq.~\eqref{ansatz_a}. Its exact benchmark tests give $\mathcal G=0$ in Minkowski spacetime and $\mathcal G=24H^4$ in de Sitter spacetime. In the GR sector it reproduces
\begin{equation}
\kappa^2\rho(0)=\frac{6M}{R^3},\qquad
\kappa^2p_r(0)=\kappa^2p_t(0)=-\frac{2(M+Rc_1)}{R^3}.
\end{equation}
For the radial conservation equation
\begin{equation}
\frac{dp_r}{dr}+\frac{a'}2(\rho+p_r)-\frac{2}{r}(p_t-p_r)=0
\label{numerical_tov}
\end{equation}
the normalized residuals are $2.5\times10^{-7}$, $3.9\times10^{-6}$, $4.7\times10^{-6}$,
and $7.1\times10^{-6}$ for $\bar\beta=0,0.5,1,3$, respectively.
The plotting script stops when this residual exceeds $10^{-3}$.

The proper thickness of the shell is defined by
\begin{equation}
\ell=\int_{r_1}^{r_2}\sqrt{e^{b(r)}}\,dr.
\end{equation}
Using $f(r)\equiv e^{-b(r)}$, we obtain the proper thickness as
\begin{equation}\label{prop}
\ell=\int_{r_1}^{r_2}\frac{dr}{\sqrt{f(r)}}.
\end{equation}

Under the thin-shell approximation,
\begin{equation}
\epsilon=r_2-r_1\ll 1,
\end{equation}
and assuming that \(f(r)\) varies slowly across the shell, we may approximate it by its value at a representative radius \(r=r_0\). Hence,
\begin{equation}
\ell \approx \frac{\epsilon}{\sqrt{f(r_0)}}.
\label{ell_approx}
\end{equation}

At fixed coordinate width $\epsilon$, a smaller $f(r_0)$ gives a
larger proper distance in Eq.~\eqref{ell_approx}. This dependence
persists when $\beta$ changes and the endpoints are held fixed.
A physical shell thickness would also depend on where its two
boundaries lie for each value of $\beta$.

Over the prescribed interval $[r_1,r_2]$, the density integral with
the coordinate weight $4\pi r^2$ is
\begin{equation}\label{47}
E=4\pi\int_{r_1}^{r_2}\rho(r)\,r^2\,dr,
\end{equation}
where $\rho(r)$ is obtained from the modified field equations.

If a sufficiently narrow shell interval is specified such that the energy
density varies slowly across it, one may approximate
\begin{equation}\label{48}
\rho(r)\approx \rho(r_0),
\end{equation}
which gives
\begin{equation}
E\approx 4\pi \rho(r_0) r_0^2 \epsilon.
\end{equation}
This coordinate integral shows that the shell contribution depends on its
thickness and radius. If one instead considers the proper matter energy,
the integrand must include the proper-volume factor $1/\sqrt{f(r)}$.

An entropy density $s\propto\sqrt{p_r}$ would require a finite shell
with a justified ultra-relativistic equation of state and $p_r\geq0$.
The profiles give neither a finite stiff-fluid shell nor a
thermodynamic law. Where $p_r<0$, the square root is non-real.

\begin{figure}[!htbp]
\centering

\begin{minipage}{0.32\textwidth}
    \centering
    \includegraphics[width=\linewidth]{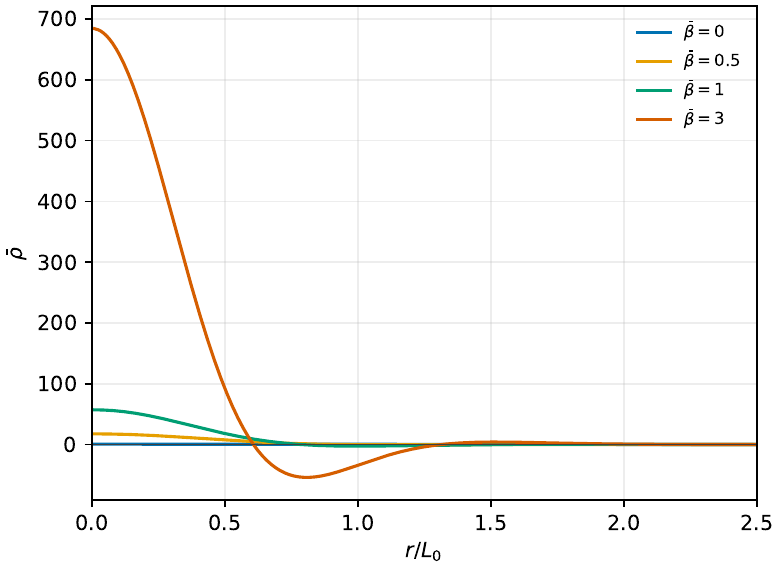}
    \textbf{(a)}
\end{minipage}
\hfill
\begin{minipage}{0.32\textwidth}
    \centering
    \includegraphics[width=\linewidth]{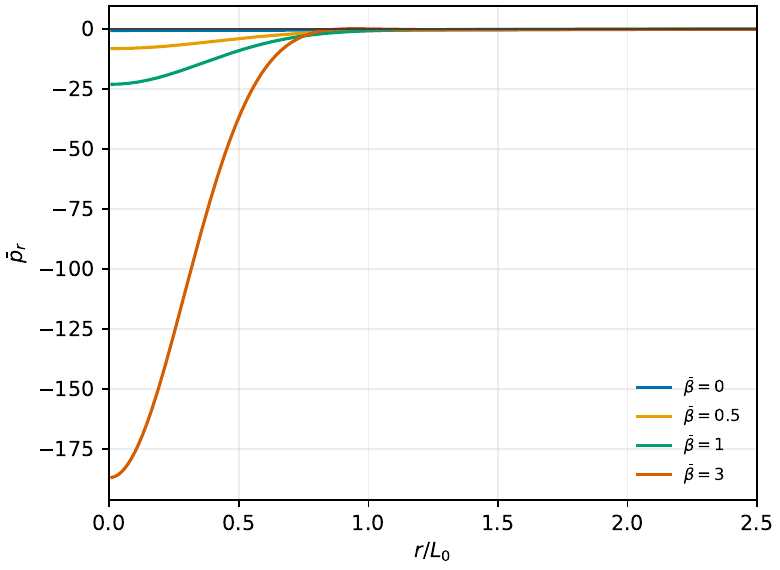}
    \textbf{(b)}
\end{minipage}
\hfill
\begin{minipage}{0.32\textwidth}
    \centering
    \includegraphics[width=\linewidth]{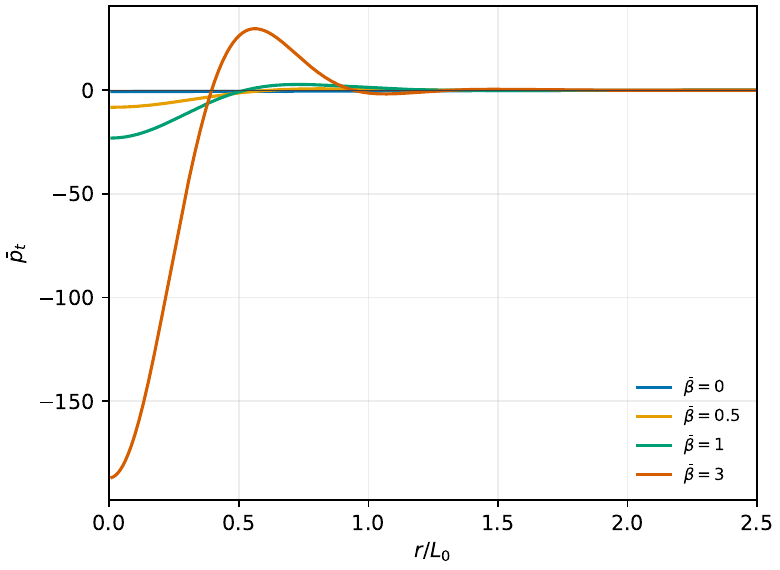}
    \textbf{(c)}
\end{minipage}

\caption{Dimensionless matter profiles obtained from the field equations of
$R+\beta\mathcal{G}^{2}$ gravity for the representative dimensionless
couplings $\bar{\beta}=0,0.5,1,3$:
(a) energy density $\bar\rho$,
(b) radial pressure $\bar p_r$, and
(c) tangential pressure $\bar p_t$, with the normalization in Eq.~\eqref{barred_matter}.}

\label{fig:matter-profiles}
\end{figure}

The matter profiles obtained from the quadratic Gauss--Bonnet field
equations are displayed in Fig.~\ref{fig:matter-profiles}. For the
representative parameter set considered here, the higher-curvature
contribution produces appreciable modifications to the required matter
distribution. For $\bar\beta>0$ the density is positive throughout the numerically
resolved near-center region, $r/L_0\geq 0.01$, but it changes sign over
intermediate radial intervals. An exact statement at $r=0$ requires
evaluation of the complete symbolic limits of $H^\mu{}_{\nu}(0)$.  The density zeros are approximately $(0.981,1.345)$ for $\bar\beta=0.5$, $(0.790,1.463)$ for $\bar\beta=1$, and $(0.607,1.309)$ for $\bar\beta=3$, in units of $L_0$. Thus these parameter choices violate the WEC in transition bands rather than at the center.

\begin{figure}[!htbp]
\centering
\includegraphics[width=0.32\textwidth]{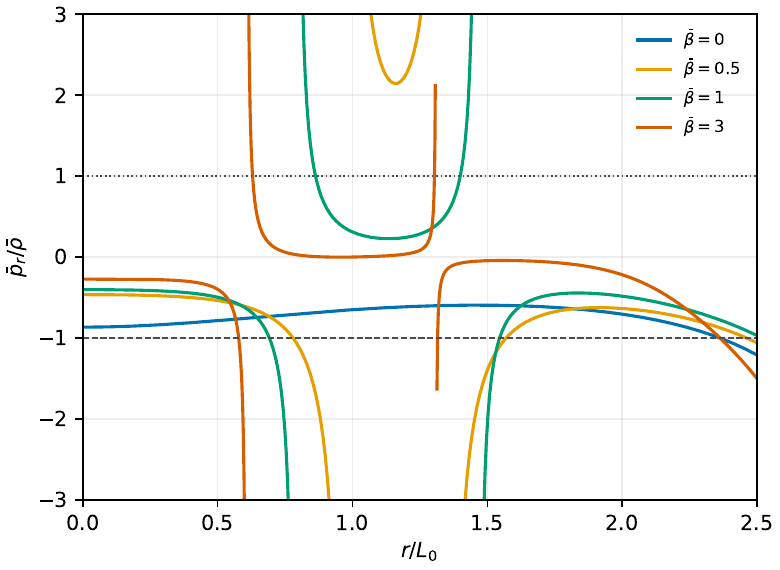}
\caption{Radial profile of $\bar p_r/\bar\rho=p_r/\rho$. Horizontal
lines at $-1$ and $+1$ mark vacuum-like and stiff-fluid behavior.
The curves are broken at zeros of $\bar\rho$, where the ratio is
undefined. Divergent branches extend beyond the displayed vertical range.}
\label{fig:eos}
\end{figure}

Figure~\ref{fig:eos} shows how $p_r/\rho$ changes through the
transition. The curves give no common finite interval over which
$p_r\simeq\rho$ holds across the parameter sets. Assigning the energy
in Eq.~\eqref{47} to a physical shell requires a criterion for the
two endpoints $r_1$ and $r_2$.

For fixed endpoints $r_1$ and $r_2$, the $\beta$ dependence of the
proper distance enters through $f(r)$,
\begin{equation}
\ell=\int_{r_1}^{r_2}\frac{dr}{\sqrt{f(r)}},
\end{equation}
so that variations in $\beta$ can modify $\ell$ through their influence on
the metric function $f(r)$. However, the physical shell boundaries are not
uniquely determined by the matter profiles considered here. Therefore,
the dependence of the physical shell thickness on $\beta$ cannot be
established without first specifying a consistent prescription for
identifying $r_1$ and $r_2$.

Similarly, the shell energy depends not only on the local matter
variables and metric functions but also on the location of the shell
boundaries. Their dependence on $\beta$ can therefore be assessed
quantitatively only after the shell region has been defined unambiguously.
For this reason, no general monotonic increase or decrease of the shell
energy with $\beta$ is inferred from the present analysis.

\section{Energy Conditions}\label{VII}

For an anisotropic fluid, the null energy condition (NEC), weak energy condition (WEC), and strong energy condition (SEC) are tested through
\begin{align}
&\mathrm{NEC}: && \rho+p_r\ge 0,\qquad \rho+p_t\ge 0,\\
&\mathrm{WEC}: && \rho\ge 0,\qquad \rho+p_r\ge 0,\qquad \rho+p_t\ge 0,\\
&\mathrm{SEC}: && \rho+p_r+2p_t\ge 0,\qquad \rho+p_r\ge 0,\qquad \rho+p_t\ge 0.
\end{align}
The anisotropy is $\Delta=p_t-p_r$.

\begin{figure}[!htbp]
\centering

\begin{minipage}{0.48\textwidth}
    \centering
    \includegraphics[width=\linewidth]{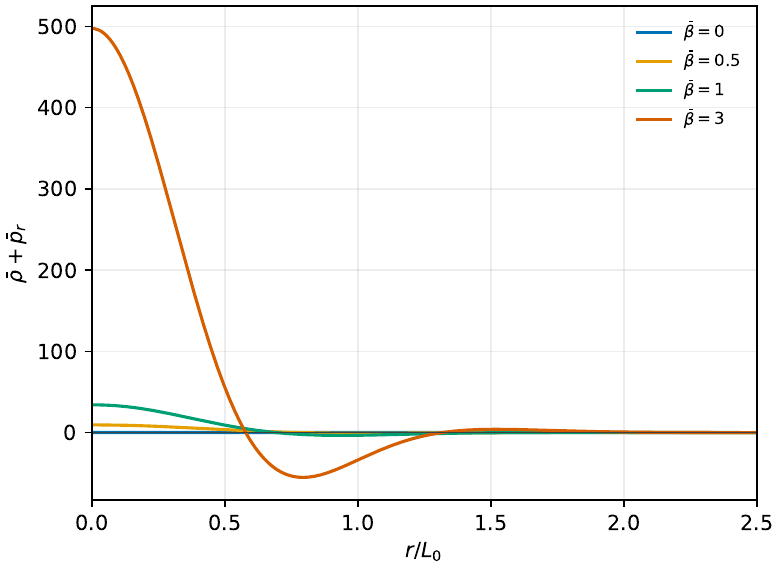}
    \textbf{(a)}
\end{minipage}
\hfill
\begin{minipage}{0.48\textwidth}
    \centering
    \includegraphics[width=\linewidth]{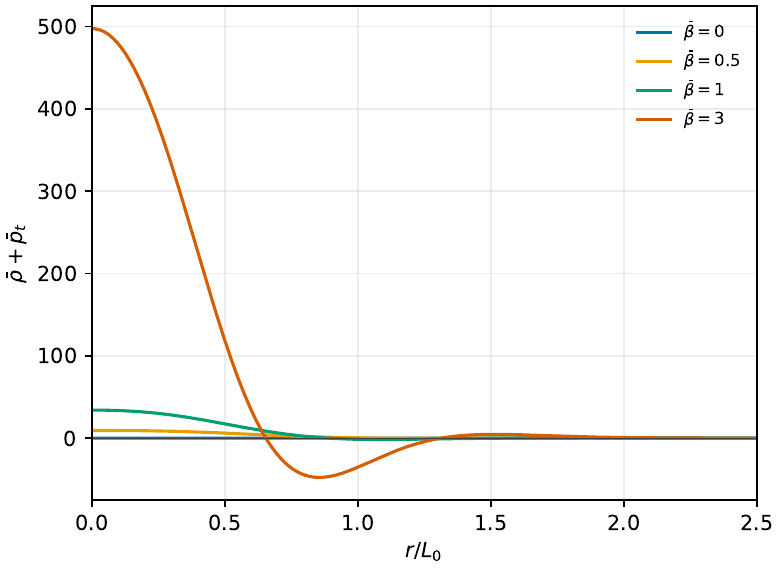}
    \textbf{(b)}
\end{minipage}

\vspace{2mm}

\begin{minipage}{0.48\textwidth}
    \centering
    \includegraphics[width=\linewidth]{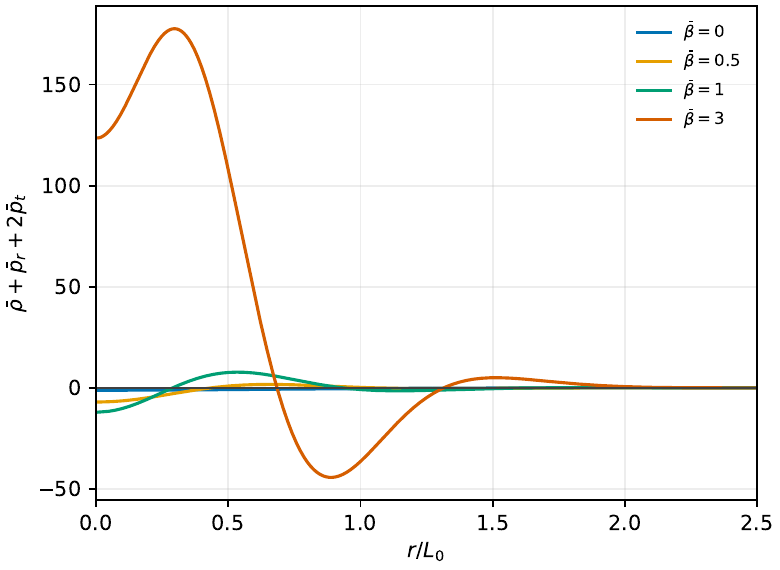}
    \textbf{(c)}
\end{minipage}
\hfill
\begin{minipage}{0.48\textwidth}
    \centering
    \includegraphics[width=\linewidth]{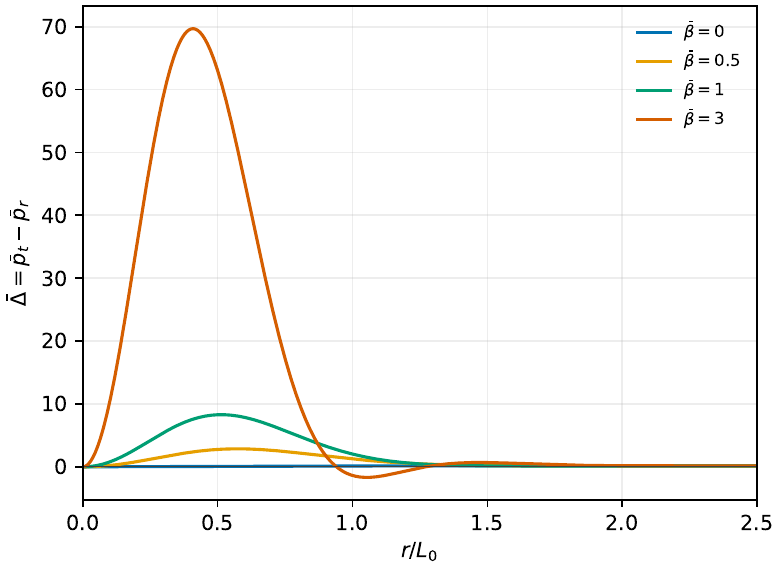}
    \textbf{(d)}
\end{minipage}

\caption{Dimensionless energy-condition combinations and pressure anisotropy for the
representative parameter values considered in this work.
(a) Radial NEC combination $\bar\rho+\bar p_r$, where positive values satisfy the
radial NEC.
(b) Tangential NEC combination $\bar\rho+\bar p_t$, where positive values satisfy
the tangential NEC.
(c) SEC combination $\bar\rho+\bar p_r+2\bar p_t$, whose sign illustrates the dependence
of the strong energy condition on the radial coordinate and the
Gauss--Bonnet coupling.
(d) Pressure anisotropy $\bar\Delta=\bar p_t-\bar p_r$.}

\label{fig:energy-conditions}
\end{figure}

Figure~\ref{fig:energy-conditions} displays the corresponding
energy-condition combinations and pressure anisotropy for the
representative dimensionless parameter set stated in Sec.~\ref{VI}. Panels (a) and (b) show the radial
and tangential NEC combinations $\rho+p_r$ and $\rho+p_t$,
respectively; panel (c) shows the SEC combination
$\rho+p_r+2p_t$; and panel (d) shows the pressure anisotropy
$\Delta=p_t-p_r$.

The energy-condition analysis reveals a pronounced dependence on both the
radial coordinate and the quadratic Gauss--Bonnet coupling. For the
representative parameter set considered here, the positive values
$\bar{\beta}=0.5,1,3$ do not lead to global satisfaction of the null and
weak energy conditions. Both NEC combinations are positive throughout the numerically resolved
near-center region for these couplings, but become negative over
intermediate radial bands.  The WEC
also fails wherever the density is negative, between the pairs of zeros
listed above. The strong energy condition changes sign with radius and is not
globally satisfied. These results demonstrate
that regularity of the geometry does not by itself guarantee satisfaction
of the standard energy conditions.

{ For the positive-mass members (\(M>0\)) of the present metric family, parameter tuning cannot restore the energy conditions globally. At large radius, the radial field equation
gives
\[
p_r=-\frac{2M}{\kappa^2 r^3}+O(r^{-4}),
\qquad
\rho+p_r=-\frac{2M}{\kappa^2 r^3}+O(r^{-4}),
\]
because the density contains no compensating positive \(r^{-3}\) term.
Furthermore, for \(\beta c_3>0\),
\[
\rho=-\frac{2\beta c_3}{\kappa^2 r^4}+O(r^{-5}),
\]
while the explicit \(G^2\) contributions decay faster asymptotically.
Thus, for \(M>0\), the radial NEC must eventually fail, and positive
\(\beta c_3\) also produces an asymptotic WEC violation. Within this
positive-mass family, varying the parameters may shift or reduce the
violating intervals on a finite radial domain, but it cannot remove the
asymptotic radial-NEC violation. A physically
viable compact configuration therefore requires matching at a finite
radius to a separately specified vacuum exterior, or adopting a
different asymptotic ansatz.}

\section{Requirements for a Stability Analysis}\label{VIII}

We now discuss an important point concerning the stability problem. A dynamical thin-shell analysis requires more information than the static junction relation alone.  In the ordinary Israel formalism one may write, for a shell radius $r=r(\tau)$,
\begin{equation}
\dot r^{\,2}+V(r)=0,
\end{equation}
with
\begin{equation}\label{80}
V(r)=\frac{f_+(r)+f_-(r)}{2}
-\left(2\pi r\,\sigma(r)\right)^2
-\frac{\left(f_+(r)-f_-(r)\right)^2}{64\pi^2r^2\sigma^2(r)}.
\end{equation}
However, the static relation
\begin{equation}
\sigma_0(r)=-\frac{1}{4\pi r}\left(\sqrt{f_+(r)}-\sqrt{f_-(r)}\right)
\end{equation}
gives the surface density of a static shell. Inserting it into
Eq.~\eqref{80} at every displaced radius makes $V(r)$ identically zero.
A radial perturbation calculation must determine how the surface
density evolves. If the net bulk momentum flux vanishes, the
surface conservation equation reduces to
\begin{equation}
\frac{d\sigma}{dr}+\frac{2}{r}(\sigma+P)=0,
\end{equation}
and can be closed with a surface equation of state or an equivalent
shell-mass prescription. In general, surface conservation also
includes the jump in the normal energy--momentum flux. With the
nonvacuum tail found here, the flux term must be evaluated from the
matched bulk solutions. Once the shell law and the two geometries
$f_-(r)$ and $f_+(r)$ are fixed, the GR potential gives the equilibrium
conditions \cite{Nashed:2021gkp}
\begin{equation}
V(r_0)=0,\qquad V'(r_0)=0,
\end{equation}
and linear stability requires
\begin{equation}\label{stb}
V''(r_0)>0.
\end{equation}

We should also note an additional issue which is specific to the present higher-curvature theory.  The action $R+\beta\mathcal{G}^2$ leads to fourth-order metric equations, so the GR Darmois--Israel conditions need not constitute the complete distributional matching conditions.  Extra continuity conditions can arise from the higher-curvature sector.  Therefore, the formulas above are retained only as the GR thin-shell limit and as a guide to the structure of a future stability calculation.  A claim of dynamical stability in the full quadratic Gauss--Bonnet theory requires a derivation of the appropriate generalized junction conditions and a consistent shell evolution law.

\section{Effects of the Parameter \texorpdfstring{$\beta$}{beta}}\label{IX}

Changing $\beta$ changes the field equations and the metric potentials in
Eq.~\eqref{ansatz_a}. The results below compare the reconstructed matter
along this family of geometries.

To see how $\beta$ affects radial distances, hold the coordinate
endpoints $r_1$ and $r_2$ fixed and use
\begin{equation}
\ell=\int_{r_1}^{r_2}\frac{dr}{\sqrt{f(r)}}.
\end{equation}
For $c_3>0$, the term proportional to $\beta$ lowers $f(r)$ as $\beta$
increases. On a fixed interval with $f>0$, the integrand increases at
each radius and so does $\ell$. To obtain the thickness of a physical
shell, one would also need the dependence of its boundary radii on
$\beta$.

The integral in Eq.~\eqref{47} samples a $\beta$-dependent density
since both the metric curvature and the field equations change. If the
interval is narrow enough for the approximation in Eq.~\eqref{48},
its coordinate energy scales as $\rho(r_0)r_0^2\epsilon$.

The Gauss--Bonnet parameter $\beta$ also enters the metric functions
that would determine the dynamics of a perturbed shell. Therefore,
$\beta$ is expected to influence the effective potential once a complete
dynamical shell prescription is specified. However, the static junction
relations alone are insufficient to determine this dependence. A
quantitative assessment of the influence of $\beta$ on dynamical stability
requires a surface equation of state or an equivalent shell-evolution law,
together with the appropriate generalized junction conditions for the
quadratic Gauss--Bonnet theory. We therefore do not infer enhancement or
suppression of stability as $\beta$ varies from the present static analysis.

From the above results, we find that the parameter $\beta$ acts as a control parameter
governing the interplay between geometry and the required matter variables
in the present model. Unlike phenomenological parameters introduced through
an equation of state, $\beta$ originates from the underlying modified
gravitational theory and directly modifies the curvature structure of the
spacetime. Its variation therefore affects the metric functions and the
resulting density, pressure, and anisotropy profiles. Its influence on
dynamical stability, however, cannot be quantified without specifying the
dynamical properties of the shell and the corresponding generalized
junction conditions.
\section{Summary and Discussion}\label{X}

In this work, within the framework of quadratic Gauss--Bonnet gravity defined by $f(R,\mathcal G)=R+\beta\mathcal G^2$, we investigated a static, spherically symmetric, and regular geometry with gravastar-like features. Our construction is geometry-driven: we first specify the metric functions and then use the gravitational field equations to determine the corresponding energy density and the radial and tangential pressures. In this way, the matter distribution is reconstructed from the geometry itself, without assuming any microscopic equation of state.

At the center $f(0)=1$ and $a'(0)=0$. The Einstein-sector expansion
gives a finite density and equal pressures. The reference GR solution
has $p(0)/\rho(0)=-13/15$, which quantifies its departure from the
de Sitter equation of state even though the leading radial metric
has de Sitter form. For $\beta\neq0$, the finite central components
of $H^\mu{}_{\nu}$ also enter the density and pressure.

At large $r$, the term $-2M/r$ appears in $f(r)$ while the first
correction to the constant temporal potential is of order $1/r^2$.
A constant rescaling of time gives an asymptotically flat metric.
The difference in these falloffs leaves a radial pressure
proportional to $r^{-3}$ even at $\beta=0$, so the asymptotic region
contains matter.

The radial metric function is positive on $0<r/L_0\leq20$ for all
four parameter sets, and $e^{a(r)}$ stays positive there. The maxima
of $\mathcal C=1-f$ are $0.6866$, $0.7225$, $0.7584$, and $0.9030$
for $\bar\beta=0,0.5,1,3$, respectively. The last case comes closest
to the radial horizon condition on this interval. The matter
calculation reproduces the Minkowski and de Sitter values of
$\mathcal G$ and the analytic GR central values. The largest
normalized radial conservation residual is $7.1\times10^{-6}$,
below the $10^{-3}$ tolerance.

The matter profiles show that the quadratic Gauss--Bonnet term can have a strong effect in the transition region. For the positive representative values $\bar\beta=0.5,1,3$, the density is positive near the center within the numerically resolved region, but becomes negative over finite intermediate intervals. The two NEC combinations also change sign, which means that the NEC and WEC are not globally satisfied for these parameter choices. The strong energy condition likewise changes sign with radius. Therefore, a regular metric does not by itself guarantee a physically acceptable matter distribution. If one aims to obtain a gravastar-like configuration that satisfies selected energy conditions in a given region, the parameters of the metric ansatz must be restricted further.

The profiles of $p_r/\rho$ provide no common finite stiff-fluid
interval for the parameter sets examined. To use the proper
thickness or density integral as a shell observable, one must supply
physical endpoints $r_1$ and $r_2$. At fixed endpoints and for
$c_3>0$, increasing $\beta$ increases the proper distance while $f$
remains positive. Any shift of the endpoints would also enter the
physical shell thickness.

The Darmois--Israel formulas give the GR reference for matching.
The $R+\beta\mathcal G^2$ field equations contain derivatives of
curvature, so joining separate interior and exterior metrics requires
the corresponding distributional junction conditions. In the GR
potential calculation, inserting the static surface density at every
displaced radius makes the potential identically zero. Testing
$V(r_0)=V'(r_0)=0$ and $V''(r_0)>0$ requires a law for how the surface
stresses or the shell mass change under a displacement.

We should further emphasize that changing $\beta$ within the present family modifies two things at once: the gravitational field equations and the metric ansatz itself. Therefore, the curves obtained for different values of $\beta$ do not describe how a single fixed geometry responds to the Gauss--Bonnet correction. Rather, they represent different members of a $\beta$-dependent family in different quadratic Gauss--Bonnet theories. This distinction is important for a correct interpretation of the numerical results.

Therefore, several further steps are required before the present configuration can be regarded as a complete gravastar model. A physical prescription must be given for the core and shell boundaries, the interior solution must be matched to a separately specified exterior using junction conditions appropriate to the nonlinear higher-curvature theory, a consistent matter law for the shell must be introduced, and a genuine perturbative stability analysis must then be carried out. These limitations also distinguish the present work from earlier $f(R,\mathcal G)$ and $f(\mathcal G)$ gravastar models, in which the three regions and their equations of state are imposed from the outset \cite{Bhatti:2020cjz,Bhatti:2021dvp}. Within its current scope, the present analysis establishes a regular, horizon-free, and highly compact geometry for the displayed parameter values, and explicitly shows how the quadratic Gauss--Bonnet sector modifies the reconstructed anisotropic matter distribution and its energy-condition behavior. It does not, however, establish a globally viable stellar matter model. In the limit $\beta\to0$, the explicit higher-curvature contribution vanishes and the general relativity sector is recovered.

The comparison in Table~\ref{I} places the present construction within the broader context of representative compact-object studies in modified gravity. We emphasize that several of the cited models describe stars or wormholes rather than gravastars. The purpose of the comparison is therefore to distinguish between gravitational frameworks and construction methods, not to suggest a one-to-one correspondence among these systems.

\begin{table*}[!t]
\caption{Scope of representative comparison studies. The distinction between
gravastar and non-gravastar constructions is made explicit to avoid a
like-for-like interpretation where none is warranted.}
\label{I}
\centering

\footnotesize
\renewcommand{\arraystretch}{1.35}
\setlength{\tabcolsep}{3pt}

\begin{tabular}{|l|l|l|l|}
\hline

\textbf{Reference / class}
&
\textbf{Theory}
&
\textbf{Construction}
&
\textbf{Relevance to the present work}
\\
\hline

Mazur \& Mottola \cite{Mazur:2004fk}
&
GR
&
Standard gravastar idealization
&
\shortstack[l]{Baseline three-region concept; distinct regions\\
and equations of state are part of the construction.}
\\
\hline

Astashenok et al. \cite{Astashenok:2013vza}
&
$f(R)$
&
Relativistic neutron-star models
&
\shortstack[l]{Modified-gravity stellar structure;\\
not a gravastar construction.}
\\
\hline

Boehmer et al. \cite{Boehmer:2011gw}
&
$f(T)$
&
Static relativistic stars
&
\shortstack[l]{Teleparallel compact-star comparison;\\
not a gravastar construction.}
\\
\hline

\shortstack[l]{Zubair et al.; Deb et al.\\
\cite{Zubair:2016cde,Deb:2022jzj}}
&
$f(R,T)$
&
Wormhole geometries
&
\shortstack[l]{Matter--geometry coupling and energy conditions;\\
not gravastar models.}
\\
\hline

Tayde et al. \cite{Tayde:2022lxd}
&
Nonmetricity gravity
&
Static wormhole geometry
&
\shortstack[l]{Higher-gravity compact geometry;\\
not a gravastar model.}
\\
\hline

\shortstack[l]{Loewer et al.; Errehymy et al.\\
\cite{Loewer:2024lai,Errehymy:2025kzj}}
&
$f(R,L_m,T)$
&
Wormhole constructions
&
\shortstack[l]{Extended matter--geometry coupling;\\
not gravastar models.}
\\
\hline

Tangphati et al. \cite{Tangphati:2020mir}
&
Massive gravity
&
Traversable wormholes
&
\shortstack[l]{Modified strong-field geometry;\\
not a gravastar construction.}
\\
\hline

\shortstack[l]{Bhatti, Yousaf \& Rehman\\
\cite{Bhatti:2020cjz}}
&
$f(R,\mathcal G)$
&
\shortstack[l]{Three-region gravastar with\\
prescribed equations of state}
&
\shortstack[l]{Direct gravastar predecessor; Israel matching,\\
shell length, entropy, and energy are evaluated.}
\\
\hline

\shortstack[l]{Bhatti, Yousaf \& Ashraf\\
\cite{Bhatti:2021dvp}}
&
$f(\mathcal G)$
&
\shortstack[l]{Three-region modified\\
Gauss--Bonnet gravastar}
&
\shortstack[l]{Direct Gauss--Bonnet comparison; isotropic matter\\
and prescribed regional equations of state.}
\\
\hline

Nojiri \& Nashed \cite{Nojiri:2023zlp}
&
\shortstack[l]{Einstein gravity with\\
constrained scalar fields}
&
\shortstack[l]{de Sitter interior, smooth shell,\\
Schwarzschild exterior}
&
\shortstack[l]{Explicit shell boundaries, large surface redshift,\\
and stability can be addressed.}
\\
\hline

Sharif et al. \cite{Sharif:2020xxj}
&
\shortstack[l]{Modified\\
Gauss--Bonnet gravity}
&
Static wormhole solutions
&
\shortstack[l]{Nonlinear Gauss--Bonnet compact geometries;\\
not a gravastar model.}
\\
\hline

\textbf{Present work}
&
$\bm{R+\beta\mathcal G^2}$
&
\shortstack[l]{\textbf{Global geometry-driven}\\
\textbf{ansatz}}
&
\shortstack[l]{\textbf{Regular geometry with reconstructed matter}\\
\textbf{and nonvacuum asymptotics; no three-region}\\
\textbf{or thin-shell stability claim.}}
\\
\hline

\end{tabular}

\end{table*}

\newpage
\appendix
\section{Near-Center Expansion and Matter Variables}\label{app:center}

In this appendix, we give the near-center expansion of the metric functions and the corresponding Einstein-sector matter variables. This verifies regularity of the metric expansion and isolates the additional conditions required for an exact de Sitter center.

We consider the static and spherically symmetric metric given by Eq.~\eqref{met}, together with the ansatz of Eq.~\eqref{ansatz_a}. Using the expansions
\[
\frac{r^2}{R^2+r^2}
= \frac{r^2}{R^2}-\frac{r^4}{R^4}+\frac{r^6}{R^6}+\mathcal O(r^8), \quad
\frac{r^4}{(R^2+r^2)^2}
= \frac{r^4}{R^4}-\frac{2r^6}{R^6}+\mathcal O(r^8),
\]
we obtain
\begin{equation}
a(r)=c_0+\alpha r^2+\eta r^4+\zeta r^6+\mathcal O(r^8),
\end{equation}
where
\[
\alpha=-\frac{c_1}{R^2}, \quad
\eta=\frac{\beta c_2+c_1}{R^4}, \quad
\zeta=-\frac{c_1+2\beta c_2}{R^6}.
\]

For the metric function $f(r)$, using
\[
1-e^{-r^3/R^3}
= \frac{r^3}{R^3}-\frac{r^6}{2R^6}+\mathcal O(r^9), \quad
\frac{r^2}{(R^2+r^2)^2}
= \frac{r^2}{R^4}-\frac{2r^4}{R^6}+\frac{3r^6}{R^8}+\mathcal O(r^8),
\]
we obtain
\begin{equation}
f(r)=1-Ar^2+Br^4+Cr^5+Dr^6+\mathcal O(r^7),
\end{equation}
with
\[
A=\frac{2M}{R^3}+\frac{2\beta c_3}{R^4}, \quad
B=\frac{4\beta c_3}{R^6}, \quad
C=\frac{M}{R^6}, \quad
D=-\frac{6\beta c_3}{R^8}.
\]

The derivatives are
\begin{align}
a'(r)&= 2\alpha r + 4\eta r^3 + 6\zeta r^5 + \mathcal O(r^7), \\
a''(r)&= 2\alpha + 12\eta r^2 + 30\zeta r^4 + \mathcal O(r^6), \\
f'(r)&= -2Ar + 4Br^3 + 5Cr^4 + 6Dr^5 + \mathcal O(r^6).
\end{align}

For the metric written in terms of $f(r)\equiv e^{-b(r)}$, the Einstein tensor components are
\begin{align}\label{ein_appendix}
-G^t_{\ t} &= \frac{1-f-rf'}{r^2}, \qquad
G^r_{\ r} = \frac{f-1}{r^2}+\frac{f a'}{r}, \\
G^\theta_{\ \theta} &= f\left(\frac{a''}{2}+\frac{a'^2}{4}-\frac{a'b'}{4}+\frac{a'-b'}{2r}\right),
\end{align}
where $b'(r)=-f'(r)/f(r)$.

Substituting the expansions yields
\begin{align}
-G^t_{\ t} &= 3A - 5Br^2 - 6Cr^3 - 7Dr^4 + \mathcal O(r^5), \\
G^r_{\ r} &= (2\alpha-A) + \big(B+4\eta-2A\alpha\big)r^2 + Cr^3 + \mathcal O(r^4), \\
G^\theta_{\ \theta} &= (2\alpha-A)
+ \big(2B+\alpha^2-3A\alpha+8\eta\big)r^2
+ \frac{5}{2}Cr^3 + \mathcal O(r^4).
\end{align}

The modified field equations are
\begin{equation}
G_{\mu\nu}+\beta H_{\mu\nu}=\kappa^2 T_{\mu\nu}, \qquad
T^\mu_{\ \nu}=\mathrm{diag}(-\rho,\,p_r,\,p_t,\,p_t),
\end{equation}
so that
\begin{align}
\kappa^2 \rho = -G^t_{\ t}-\beta H^t_{\ t}, \quad
\kappa^2 p_r = G^r_{\ r}+\beta H^r_{\ r}, \quad
\kappa^2 p_t = G^\theta_{\ \theta}+\beta H^\theta_{\ \theta}.
\end{align}

At the center ($r=0$), the spacetime is regular, with
\[
a(r)=a_0+\alpha r^2+\cdots, \qquad f(r)=1-Ar^2+\cdots,
\]
and curvature invariants remain finite.

In evaluating the central limit, the contribution arising from the
quadratic Gauss--Bonnet term $\mathcal{G}^2$ must be retained in general.
It may be omitted only when its vanishing at $r=0$ follows explicitly
from the adopted metric functions and the corresponding field equations. The central matter variables therefore read
\begin{align}
\kappa^2\rho(0)&=3A-\beta H^t{}_{t}(0),\\
\kappa^2p_r(0)&=2\alpha-A+\beta H^r{}_{r}(0),\\
\kappa^2p_t(0)&=2\alpha-A+\beta H^\theta{}_{\theta}(0).
\end{align}
Regular spherical symmetry requires equality of the two central pressures, but the finite higher-curvature terms need not vanish.  Hence the earlier GR-like central formulas and the condition $\alpha=-A$ are not assumed.  If an exact de Sitter core is required, this condition must be verified
directly from the field equations by imposing
\begin{equation}
p_r(0)=p_t(0)=-\rho(0).
\end{equation}
For the numerical parameters used in this manuscript we therefore claim only a regular, gravastar-like core unless this equality is verified explicitly after recomputation of $H^\mu{}_{\nu}$.

\section{Quadratic Gauss--Bonnet Tensor}\label{app:H}

For the quadratic Gauss--Bonnet model
\begin{equation}
f(R,\mathcal{G})=R+\beta\mathcal{G}^{2},
\end{equation}
the field equations may be written in the form
\begin{equation}
G_{\mu\nu}+\beta H_{\mu\nu}
=\kappa^{2}T_{\mu\nu},
\end{equation}
where $H_{\mu\nu}$ denotes the contribution associated with the
quadratic Gauss--Bonnet term. After factoring out the coupling
$\beta$, and using the convention in Eq.~\eqref{riemann_convention}, it is given by
\begin{align}
H_{\mu\nu}={}&
4\mathcal{G}R R_{\mu\nu}
-8\mathcal{G}R_{\mu\alpha}R^{\alpha}{}_{\nu}
-8\mathcal{G}R_{\mu\alpha\nu\beta}R^{\alpha\beta}
+4\mathcal{G}R_{\mu}{}^{\alpha\beta\gamma}
R_{\nu\alpha\beta\gamma}
-\frac{1}{2}g_{\mu\nu}\mathcal{G}^{2}
\nonumber\\
&-4R\nabla_{\mu}\nabla_{\nu}\mathcal{G}
+4g_{\mu\nu}R\Box\mathcal{G}
+8R^{\alpha}{}_{\nu}\nabla_{\alpha}\nabla_{\mu}\mathcal{G}
+8R^{\alpha}{}_{\mu}\nabla_{\alpha}\nabla_{\nu}\mathcal{G}
\nonumber\\
&-8R_{\mu\nu}\Box\mathcal{G}
-8g_{\mu\nu}R^{\alpha\beta}
\nabla_{\alpha}\nabla_{\beta}\mathcal{G}
+8R_{\mu\alpha\nu\beta}
\nabla^{\alpha}\nabla^{\beta}\mathcal{G}.
\end{align}

The numerical coefficients in $H_{\mu\nu}$ are obtained from
\begin{equation}
f_{\mathcal{G}}
\equiv
\frac{\partial f}{\partial\mathcal{G}}
=2\beta\mathcal{G}.
\end{equation}
Therefore, the quadratic Gauss--Bonnet contribution depends on both
the curvature invariant $\mathcal{G}$ and its covariant derivatives.
For the static and spherically symmetric geometry considered in this
work, the corresponding $tt$, $rr$, and angular components determine the required matter variables $\rho(r)$, $p_r(r)$, and  $p_t(r)$, respectively.

\section*{Computational Reproducibility}

The complete symbolic derivation, benchmark tests, numerical evaluation,
conservation diagnostic, density-zero detection, and figure-generation
procedure should accompany the manuscript as the file
\texttt{quadratic\_gauss\_bonnet\_gravastar.py}. The program requires
SymPy, NumPy, SciPy, and Matplotlib. Because the numerical curves depend on this implementation, the code is part of the reproducibility record rather than an optional supplement.

The implementation constructs the Christoffel symbols, Riemann tensor,
Ricci tensor, Ricci scalar, Einstein tensor, Gauss--Bonnet invariant, and
quadratic Gauss--Bonnet tensor covariantly using the curvature convention
in Eq.~\eqref{riemann_convention}. It verifies
\[
\mathcal G_{\mathrm{Minkowski}}=0,
\qquad
\mathcal G_{\mathrm{dS}}=24H^4,
\]
and reproduces the analytic GR central limits. The numerical plotting
routine also evaluates the conservation residual
\[
\mathcal R_{\mathrm{TOV}}=
\frac{dp_r}{dr}
+\frac{a'}{2}(\rho+p_r)
-\frac{2}{r}(p_t-p_r).
\]
Figure generation is terminated automatically if the normalized residual
exceeds $10^{-3}$. For the parameter values used in this work, the measured
normalized residuals range from $2.5\times10^{-7}$ to
$7.1\times10^{-6}$.
\section*{Data Availability}
No external observational data were used in this study. The Python source
code required to reproduce the symbolic checks, numerical results, and
figures is provided as supplementary material.
%

\end{document}